\documentclass{aa}  

\usepackage{graphicx}
\usepackage{subcaption}
\usepackage{balance}
\usepackage{txfonts}
\usepackage{amsmath}
\usepackage{amssymb}
\usepackage{adjustbox}
\usepackage[colorlinks=true,linkcolor=blue,citecolor=blue,urlcolor=blue]{hyperref}
\newcommand{\lya}{Ly{$\rm \alpha$}\xspace}
\newcommand{\jwst}{{\em JWST}\xspace}
\newcommand{\hi}{\ion{H}{i}}

\newcommand{\oiii}{\ion{O}{iii}}
\newcommand{\oii}{\ion{O}{ii}}

\begin{document}

    \title{All the Massive Galaxy Overdensities during Reionization}
    \subtitle{\jwst\ Rest-Frame Optical Selection Reveals Young, Chemically Evolved Galaxies Embedded in Dense, Neutral Gas at $z>5$}

   \author{Chamilla Terp\inst{1, 2}
          \and
          Kasper E. Heintz\inst{3, 1, 2}
          \and
          Jorryt Matthee\inst{4}
          \and
          Rohan P. Naidu\inst{5}
          \and
          Pascal A. Oesch\inst{6, 1, 2}
          \and
          Callum Witten\inst{6}
          \and 
          Daichi Kashino\inst{7}
          \and
          Clara L. Pollock\inst{1, 2}
          \and
          Claudia Di Cesare\inst{4, 8}
          \and 
          Alberto Torralba\inst{9, 10}
          }

   \institute{Cosmic Dawn Center (DAWN), Denmark
            \and
             Niels Bohr Institute, University of Copenhagen, Jagtvej 128, 2200 Copenhagen N, Denmark
             \and 
             DTU Space, Technical University of Denmark, Elektrovej 327, DK2800 Kgs. Lyngby, Denmark
             \and 
             Institute of Science and Technology Austria (ISTA), Am Campus 1, 3400 Klosterneuburg, Austria
             \and
             MIT Kavli Institute for Astrophysics and Space Research, 70 Vassar Street, Cambridge, MA 02139, USA
             \and
             Department of Astronomy, University of Geneva, Chemin Pegasi 51, 1290 Versoix, Switzerland
             \and
             National Astronomical Observatory of Japan, 2-21-1 Osawa, Mitaka, Tokyo 181-8588, Japan
             \and
             INAF/Osservatorio Astronomico di Roma, Via di Frascati 33, 00078 Monte Porzio Catone, Italy
             \and
             Observatori Astronòmic de la Universitat de València, Ed. Instituts d’Investigació, Parc Científic. C/ Catedrático José Beltrán, n2, 46980 Paterna, Valencia, Spain
             \and
             Departament d’Astronomia i Astrofísica, Universitat de València, 46100 Burjassot, Spain
             }

   \date{Received XX; accepted XX}

 
  \abstract
    {The high-redshift progenitors of present-day galaxy clusters are believed to substantially contribute to the global star-formation rate density and drive the large-scale reionization of the Universe. Here we present a blind and unbiased search for and characterization of galaxy overdensities during the reionization epoch at redshifts $z\sim 5.5-7$, based on rest-frame optical \jwst/NIRCam grism spectroscopy of the Abell\,2744 lensing field as part of the \jwst-ALT survey. Using a physically-motivated, cosmological inference Friends-of-Friends (FoF) algorithm, we identify six galaxy overdensities, including five robust systems at $z=5.66$ to $6.77$. They are all characterized by total halo masses $M_{\rm halo} \gtrsim 10^{11}\,M_{\odot}$ inferred from a range of proxies. We find that the galaxy members in these overdense environments are on average less massive though equally metal-rich, and generally comprised of younger stellar populations as indicated from their bluer spectral slopes less prominent Balmer breaks, than field galaxies at similar redshifts. This is further supported by their systematically bluer UV continua and weaker Balmer breaks, indicating younger luminosity-weighted stellar populations and a reduced contribution from older stars. Further, we use this novel rest-frame optical selection of galaxy proto-clusters to infer the fraction and 3D distribution of strong Lyman-$\alpha$ emitters (LAEs) and damped Lyman-$\alpha$ absorbers (DLAs) in the overdensity environments. We find that two out of six galaxy overdensities have excess \hi\ absorption compared to the field-average, while the other four are consistent within their large scatter in density. These results present the first direct observational constraints on the tomography of the dense, neutral gas reservoirs in large-scale galaxy overdensities at $z>5$ and highlight the limitations of pre-\jwst searches for reionization-era galaxy overdensities relying on the detection of strong LAEs alone.}


   \titlerunning{All the Massive Galaxy Overdensities during Reionization}
   \authorrunning{Terp et al.}

   \maketitle
%

\section{Introduction}
Galaxy clusters are the titans of cosmic structure -- immense, gravitationally bound systems that act as key drivers of galaxy evolution and large-scale structure formation, and are powerful probes of dark matter distribution of the Universe. Their progenitors, high-redshift `proto-clusters', are particularly important for understanding how galaxies assembled in the early Universe \citep{Overzier_2016}. In these dense environments, galaxies are expected to undergo accelerated growth, fueled by cold gas inflows and mergers, and to contribute significantly to the reionization of the intergalactic medium (IGM) through their enhanced production of ionizing photons \citep{Chiang_2017, Daddi_2022}. Studying such systems therefore provides critical insights into how galaxies and the large-scale structure co-evolved in the first billion years after the Big Bang. 

A central aspect of this picture is the role of neutral atomic hydrogen (\hi) gas. As the primary fuel for star formation, cold gas accretion onto proto-galactic dark matter haloes plays a central role in regulating galaxy growth, particularly in protocluster environments where accretion rates are expected to be high \citep[e.g.,][]{Keres_2005,Dekel_2009,Schaye_2010}. Cold gas streams feeding the densest nodes of the cosmic web suggest that substantial reservoirs of \hi\ may be present in and around forming clusters, simultaneously driving star formation and tracing the buildup of the circumcluster medium (CCM). Beyond their influence on galaxy growth, these reservoirs also shape the ionization state of the surrounding IGM. At $z \sim 6$, the Universe approaches the end of the reionization epoch -- though some studies have favored a later end to reionization (e.g., \citealt{Keating_2019}; \citealt{Bosman_2022}). In this regime, overdense regions may host locally enhanced reservoirs of neutral gas, where the neutral hydrogen fraction remains elevated due to a combination of delayed ionization, self-shielding, and continuous replenishment by cold inflows. Observational evidence already points to significant neutral gas in these environments, with column densities reaching $N_{{\rm HI}} \gtrsim  10^{22} \; \rm cm^{-2}$ \citep{Terp_2024, Witten_2025_7_66_cluster, Heintz_2026_Nat}. In addition, early dust enrichment in overdense regions may further suppress the escape of ionizing photons, thereby helping to maintain a high neutral fraction  (e.g., \citealt{Hashimoto_2023}).
Within the $\Lambda$CDM framework, these systems naturally arise from the inhomogeneous dark matter (DM) distribution: small-scale DM haloes collapse to form the first galaxies, while larger-scale ($\sim$ 0.1--1 pMpc) overdensities grow into more loosely bound galaxy associations that later virialise into clusters. When these haloes reach a transitional mass of $\log_{10}(M_{\rm halo}/M_\odot) \sim 11.8$ \citep{Dekel_2006}, infalling gas becomes shock-heated, creating a hot circumgalactic medium (CGM) that suppresses star formation. This transition corresponds to the halo virial temperature exceeding the regime of efficient radiative cooling, allowing stable virial shocks to form and prevent rapid gas accretion. At higher redshifts, however, cold streams can still penetrate this hot phase \citep{Katz_N_2003, Keres_2005, Mandelker_2016, Bennett_2020}, leading to the expectation that proto-clusters should host compact, intensely star-forming galaxies that in total dominate the cosmic star-formation-rate density \citep{Behroozi_2013, Chiang_2017, Morokuma-Matsui_2025}. 


However, emerging observations reveal a more complex reality. While some proto-cluster galaxies appear extremely evolved -- with high metallicities, older stellar populations, or strong Lyman-$\alpha$ damping wings, so-called damped Lyman-$\alpha$ absorbers \citep[DLAs;][]{Wolfe_2005}, from dense \hi\ reservoirs ($N_{\hi} > 10^{22}{\rm cm^{-2}}$; \citealp{Morishita_2025,Li_Q_2025_a, Witten_2025_7_88_cluster,Wang_2025}) -- others show much younger, more typical high-$z$ properties, with no clear signs of accelerated evolution \citep{Laporte_2022, Li_Q_2025_b, Helton_2024}. This diversity suggests that environmental effects vary substantially across proto-clusters, likely reflecting differences in assembly histories, accretion rates, halo masses and ionization environments \citep[see e.g.,][]{Witten_2025_7_66_cluster}. To understand what drives these variations, a larger and more diverse sample of well-characterized overdensities is required.


The advent of the \textit{James Webb Space Telescope} (\jwst) now makes this possible. Its sensitivity and near-infrared spectroscopic capabilities allow direct detection of strong Lyman-$\alpha$ absorption in galaxies, revealing galaxy DLAs with $N_{\hi} \geq 10^{22}{\rm cm^{-2}}$ at $z>8$ \citep{Heintz_2024_exDLA}, and its rest-frame optical selection mitigates biases inherent to earlier \lya- or UV-based cluster searches \citep[e.g.][]{Castellano_2018, Leonova_2022}. \jwst\ enables the detection of both massive and low-mass protocluster members \citep{Helton_2024, Helton_2024a, fudamoto_2025}, providing a direct census of cold gas across entire structures during the epoch of reionization (EoR).

Here, we present a blind, rest-frame optical search for galaxy overdensities in the EoR at $z\gtrsim 5$, using the slitless \jwst/NIRCam grism spectroscopic observations of the Abell\,2744 lensing field obtained as part of the `All the Little Things' (ALT) survey \citep[prog. ID: 3516, PIs: Matthee \& Naidu;][]{ALT_paper_2024}. The main goal is to characterize the stellar populations and chemical enrichment of the cluster members and determine the \hi\ gas tomography of these massive, large-scale overdensities. We have structured the paper as follows. In Sect.~\ref{sec:obs}, we detail the \jwst-ALT and archival observations of the targeted field. In Sect.~\ref{sec:ODs}, we present our search for massive galaxy overdensities at $z>5$ and characterize their general properties. In Sect.~\ref{sec:gal_properties}, we outline the characterization of the physical properties of each cluster members, compared also to typical field galaxies and in Sect.~\ref{sec:neutral_gas} we construct the \hi\ absorption tomography and derive the fractions of strong LAEs and DLAs in each identified proto-cluster. Finally in Sect.~\ref{sec:conc}, we discuss and conclude on our work. Throughout the paper, we assume the concordance flat $\Lambda$CDM cosmology, with $H_0 = 67.7$\,km\,s$^{-1}$\,Mpc$^{-1}$, $\Omega_{\rm m} = 0.310$, and $\Omega_{\Lambda} = 0.689$ \citep{Planck18}.

\section{Observations}\label{sec:obs}

\subsection{ALT JWST/NIRCam grism spectroscopy}
In this work, we utilize  NIRCam imaging data from the \textit{`All the Little Things'} (ALT) survey, conducted as a part of \jwst Cycle 2 (Prog. ID 3516, PIs: Matthee \& Naidu). Briefly, ALT is designed to investigate the physical properties and environments of galaxies during the EoR, combining deep imagining and wide-field slitless spectroscopy in the well-studied Abell\,2744 lensing field \citep{ALT_paper_2024}. The survey delivers ultra-deep imagining in the F070W and F090W filters, reaching observed depths of $\sim 30 \rm \; mag$ ($5\sigma$) over a $30 \rm \; arcmin^{2}$ field, enabling the selection of extremely faint high-redshift sources at $z > 6$. 

In addition to imaging, ALT provides deep NIRCam wide-field slitless spectroscopy in the F356W filter, optimized to capture the [\oiii]\,$\lambda\lambda 4960,5008$ doublet and the H$\beta$ emission lines at $z\simeq6-7$. At higher redshifts ($z \gtrsim 7$), these lines redshift out of the F356W bandpass, and galaxies are instead identified primarily through H$\gamma$ emission and, where available, additional rest-frame optical features \citep[see][]{ALT_paper_2024}. The survey uses a dual–roll angle ``butterfly'' mosaic to minimize contamination, yielding deep spectra and secure redshifts for hundreds of galaxies at the target redshifts. We here only consider sources with robust spectroscopic line redshifts and identifications, requiring a signal-to-noise (S/N) of $>5$ for [\oiii]\,$\lambda 5007$. Source photometry and spectral energy distribution (SED) fitting, described in detail in \cite{ALT_paper_2024}, provide estimates of the stellar mass, star-formation rate (SFR), rest-frame UV slopes ($\beta_{\rm UV}$), and other physical parameters for each identified galaxy. These data products lay the foundation for the rest-frame optical identification and characterization of high-redshift galaxy overdensities and for subsequent analysis of their neutral gas content presented in this work.



\subsection{Auxillary \jwst/NIRSpec spectra and NIRCam imaing}

To complement the ALT dataset, we also query the DAWN \jwst\ Archive (DJA)\footnote{\url{https://dawn-cph.github.io/dja/}}, a homogeneously processed compilation of all publicly available \jwst\ data \citep[see e.g.,][for the technical details]{Heintz_2025,DeGraaff_2025,Pollock_2025} for existing \jwst/NIRSpec Prism spectroscopy ($\mathcal{R}\sim 100$, $\lambda = 0.6-5.5\mu$m) of the identified targets. These were obtained primarily as part of the {\tt UNCOVER} survey \citep[Labbé \& Bezanson, JWST-GO\#2561][]{Bezanson_2024}. We will discuss these axillary spectra and their derived properties more in Sect.~\ref{sec:neutral_gas}. 

To complement the \jwst/NIRCam imagining in the F070W and F090W filters delivered by {\tt ALT}, we also include photometric data from the {\tt MegaScience} survey \citep{Suess22}, providing images in all the NIRCam medium-bands of the Abell\,2744 lensing field, and additional data from the NIRCam parallel programs \#2756, PI: Chen \citep[e.g.,][]{Chen22} and \#3990, PI: Morishita. Across the {\tt ALT} footprint, imaging is thus available for in all the 20 NIRCam medium and broad bands as described in \citet{ALT_paper_2024}.

\section{Analysis and results}

\subsection{Identifications of galaxy overdensities at $z>5$} \label{sec:ODs}
Galaxy overdensities are typically identified using statistical techniques that analyze the spatial distribution of galaxies to detect significant peaks in density above the cosmic average \citep[e.g.,][]{Calvi_2021}. A wide range of clustering algorithms exist, each with subtle methodological differences \citep[see e.g.,][]{Overzier_2016}. One key distinction lies in the scale of association they consider -- either galaxy-to-galaxy or halo-to-galaxy links. Here, we employ a galaxy-to-galaxy approach using a Friends-of-Friends (FoF) algorithm, as halo-based linking has been shown to be less effective at identifying low-mass groups \citep{Robotham_2011}. The standard FoF algorithm identifies galaxy groups by linking galaxies that are within a predefined separation, defined as the linking length. This approach assumes that galaxies within the same overdense structure are more closely spaced than those in the surrounding field. Its lack of symmetry assumptions makes it well-suited for identifying irregular, filamentary, and evolving structures, as expected for galaxy overdensities and proto-clusters in the early Universe. The linking length is commonly defined as 
\begin{equation}
    l = b \times n^{-\frac{1}{3}},
    \label{eq:linking}
\end{equation}
where $n$ is the galaxy number density and $b$ is a dimensionless scaling factor controlling the strictness of group associations. Larger $b$ values link more loosely associated galaxies, while smaller values favor the identification of compact, tightly bound structures. Although a canonical value of $b = 0.2$ is commonly used for virialized systems \citep{More_2011}, this is not appropriate at high redshift, where overdensities are expected to be dynamically young and unvirialized. We therefore adopt a larger linking length, $b = 1$, which is better suited for identifying extended proto-cluster structures.

\begin{figure*}
    \centering
    \includegraphics[width=18cm]{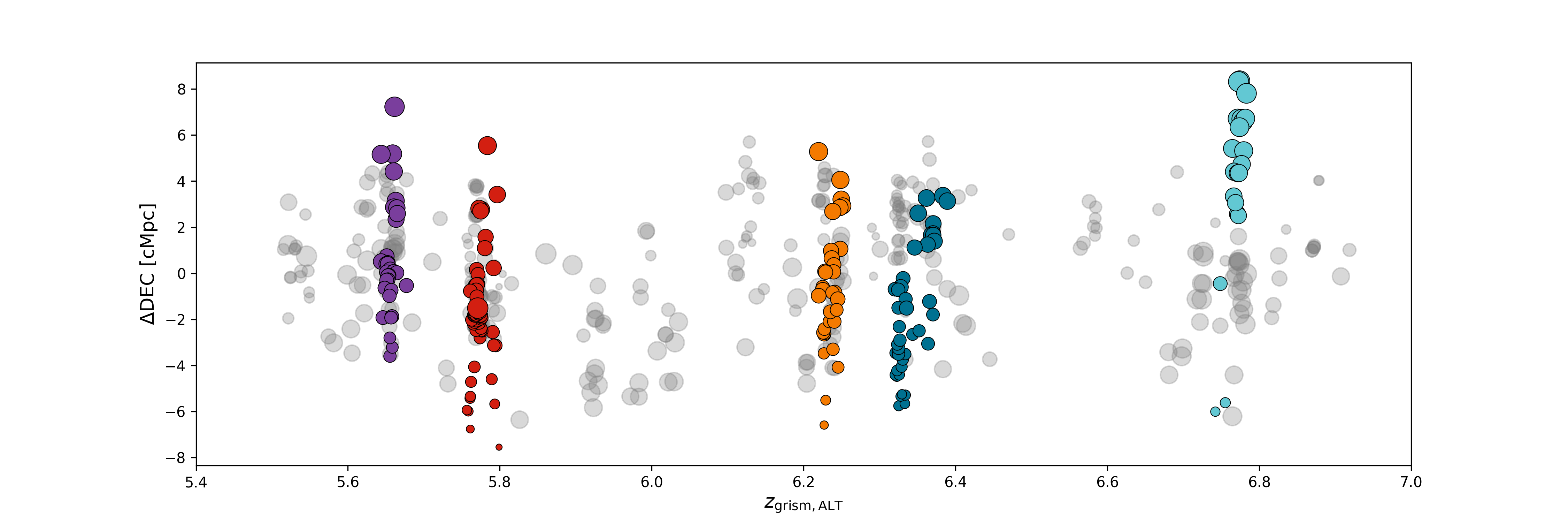}
    \caption{Spatial and redshift distribution of galaxies in the ALT catalog with overdensities marked. Spatial and redshift distribution of galaxies from the ALT catalog in the range $5.5 < z < 7$. The horizontal axis shows the grism redshift ($z_{\rm grism, ALT}$), while the vertical axis indicates the comoving offset in declination ($\rm \Delta DEC ; [cMpc]$). Marker size reflects the comoving offset in right ascension ($\rm \Delta RA ; [cMpc]$), ranging from $-6.5$ to $5.3$ cMpc, with larger symbols corresponding to larger offsets. Galaxies belonging to the five most significant overdensities identified by the FoF algorithm are shown in color, while all field galaxies are displayed in gray.}
    \label{fig:ODs_introfig}
\end{figure*}

To ensure that the linking length reflects the underlying galaxy distribution in the ALT catalog, a physically motivated, redshift-dependent value is computed. This is calculated from the comoving volume per unit solid angle within a redshift interval: 
\begin{equation}
    \frac{dV_{C}}{dz  d\Omega} = D_{H}\frac{(1+z)^{2}D_{A}^{2}}{E(z)}, \label{eq:comoving_volume}
\end{equation}
where $D_{H} = c/H_{0}$ is the Hubble distance, $D_{A}$ is the angular diameter distance, and $E(z) = H(z)/H_{0}$ is the dimensionless Hubble parameter. Integrating over a redshift interval $[z_{\text{min}}, z_{\text{max}}]$ yields the total comoving volume,
\begin{equation} 
V_{\text{full}} = 4\pi \int_{z_{\text{min}}}^{z_{\text{max}}} \frac{dV_{C}}{dzd\Omega} dz. 
\end{equation}
The effective volume covered by the survey is then 
\begin{equation} 
V = f_{\text{sky}} \times V_{\text{full}}, 
\end{equation}
where $f_{sky}$ is the fractional sky coverage of the ALT survey. The galaxy number density is then simply given by $n = N_{\text{galaxies}}/V$, where $N_{\text{galaxies}}$ is the number of sources within the survey volume in the specified redshift range. 

The FoF algorithm is applied to then group galaxies based on their 3D spatial positions in the source plane \citep[using the lensing model from][]{Furtak_2023_lensingmodel} and redshift proximity. This is done in six redshift intervals (to reduce computing time), representing the largest overdensities of sources identified visually; $z = [5.50, 5.69]$, $[5.69, 5.80]$, $[6.18, 6.30]$, $[6.30, 6.50]$, and $[6.50, 7.00]$. For each interval, a redshift-dependent linking length is computed using Eq. \ref{eq:linking}, based on the corresponding number density. The grouping is performed using \texttt{SciPy}'s \texttt{cKDTree} for efficient neighbor searches, linking galaxies that lie within the linking length of each other. 

The resulting physically-motivated linking lengths, number densities, and survey volumes for each redshift interval are summarized in Table \ref{table:linking_lengths}. In contrast to previous \jwst\ studies that adopt a fixed comoving linking length \citep[e.g.,][]{Helton_2024, Helton_2024a}, we allow the linking length to vary with redshift according to the evolving galaxy number density. This approach leads to the identification of fewer, but systematically richer, overdensities.

For comparison, applying a fixed linking length of 500 pkpc (corresponding to $\sim3.5$ cMpc at $z\sim6$) results in a larger number of groups, but with lower typical membership. Using the physically motivated, redshift-dependent linking lengths instead yields overdensities with higher average galaxy membership, consistent with the identification of more extended or filamentary structures compared to those selected using a fixed linking length. The overall population of prominent overdensities, however, remains qualitatively similar between the two approaches.

\begin{table}[h!]
\centering
\caption{Computed physically motivated linking lengths ($l$), number densities ($n$) and effective survey volumes ($V$) for each redshift range.}
\begin{adjustbox}{width=9cm}
\begin{tabular}{c c c c} 
 \hline
  Redshift range & $l$ [cMpc] & $n$ [galaxies cMpc$^{-3}$] & $V$ [cMpc$^{3}$] \\ [0.5ex] 
 \hline\hline
$[5.50, 5.69]$ & 6.1 & 4.36 $\times 10^{-3}$ & 1.54 $\times 10^{4}$ \\
$[5.69, 5.80]$ & 5.1 & 7.20 $\times 10^{-3}$ & 8.75 $\times 10^{3}$ \\
$[6.18, 6.30]$ & 5.9 & 4.87 $\times 10^{-3}$ & 9.03 $\times 10^{3}$ \\
$[6.30, 6.50]$ & 6.6 & 3.45 $\times 10^{-3}$ & 1.48 $\times 10^{4}$ \\
$[6.50, 7.00]$ & 8.3 & 1.74 $\times 10^{-3}$ & 3.56 $\times 10^{4}$ \\
 [1ex] 
\hline
\end{tabular}
\end{adjustbox}
\label{table:linking_lengths}
\end{table}
We select robust high-redshift galaxy overdensities from the groups identified by the FoF algorithm by imposing a minimum membership criterion of $N_{\mathrm{gal}} \geq 5$. Applying this threshold yields 11 overdensities out of 84 FoF-identified groups, with galaxy membership counts of $N_{\mathrm{gal}} = 8-59$. These structures are found to be largely robust to variations in the linking parameter $b$, with similar groupings recovered even when $b$ is varied by factors of a few. The subsequent analysis focuses on the five largest overdensities, located at $z \sim$ 5.66, 5.77, 6.24, 6.34, and 6.77 as shown in Figure \ref{fig:ODs_introfig}. We also identify an overdensity at $z = 7.88$, obtained by computing the linking length within the redshift interval $7.00 < z < 8.00$. This structure has been previously studied in several works \citep[see e.g.,][]{Ishigaki_2016, Hashimoto_2023, Morishita_2025b, Witten_2025_7_88_cluster}. However, we exclude this overdensity from parts of the analysis in Sect. \ref{sec:gal_properties} because, at $z > 7$, only H$\gamma$ and \oii\; are accessible, whereas the $5 < z < 7$ sample is characterized using \oiii\; and H$\beta$ \citep{ALT_paper_2024}. This leads to non-uniform selection and prevents a direct comparison.

The significance of each overdensity is quantified using the galaxy overdensity parameter, $\rm \delta_{gal} = \textit{n}_{group} / \overline{\textit{n}} - 1$, where $\rm \textit{n}_{group}$ is the number density of galaxies within the group, and $\rm \overline{\textit{n}}$ is the mean number density of galaxies in the corresponding redshift range. The group density is estimated by dividing the number of member galaxies by the volume of the convex hull enclosing their spatial distribution. To prevent artificially large overdensities arising from very small group volumes, a minimum volume threshold of $V_{min} = 161.3 \; \mathrm{cMpc}^{3}$ is imposed (calculated from the whole survey), defined as a fraction of the mean volume per galaxy in the survey. The mean density $\rm \overline{\textit{n}}$ is computed from the full galaxy sample within each redshift bin using the total survey volume.

\begin{figure*}[h!]
    \centering
    \includegraphics[width=18cm]{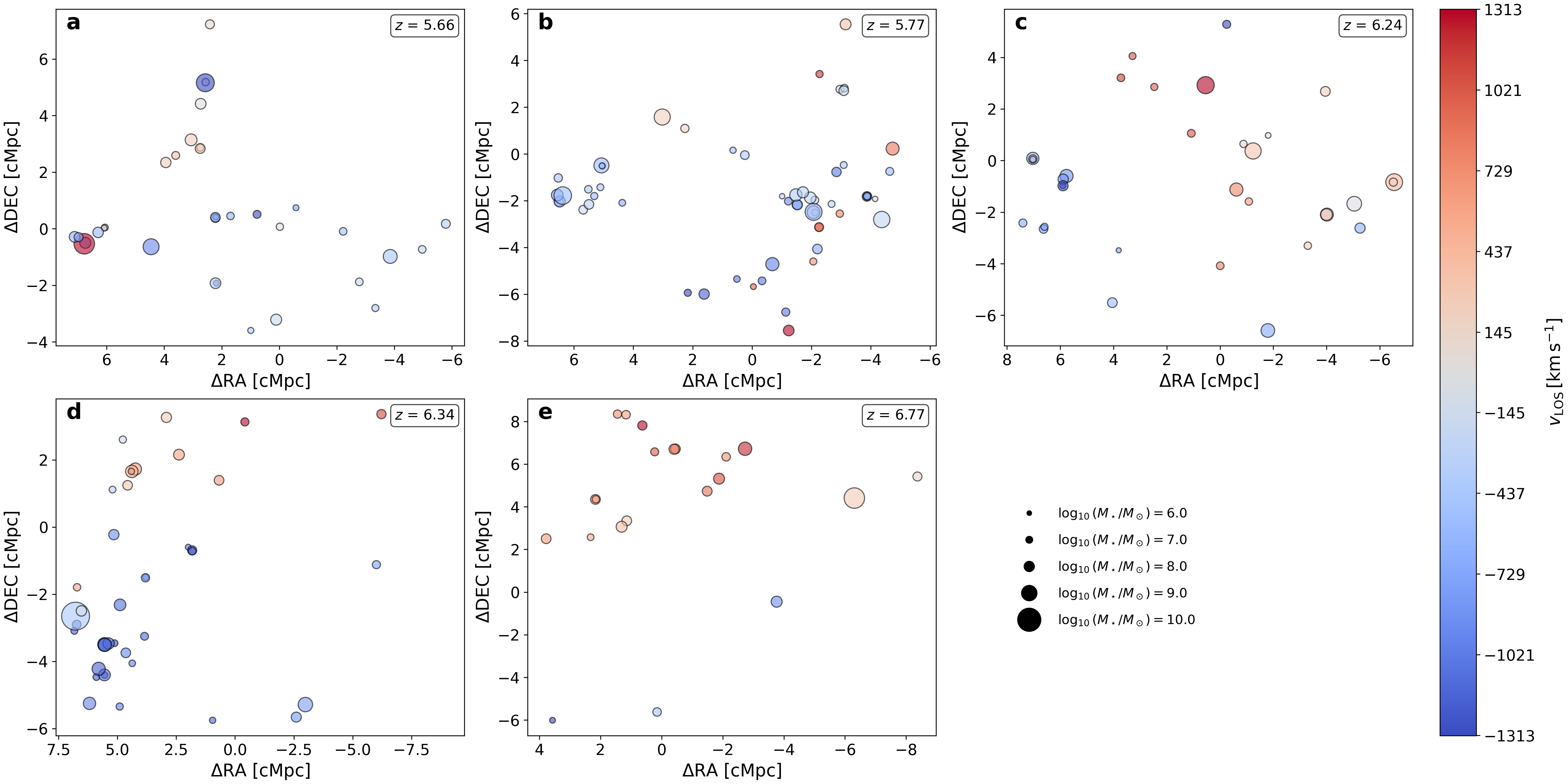}
    \caption[Projected spatial and velocity structure of overdensities]{Projected spatial distribution of galaxies in the five largest overdensities identified in the ALT survey, plotted in comoving coordinates (RA vs. DEC). Each panel corresponds to a different overdensity, labeled by its mean redshift: (a) $z = 5.66$, (b) $z = 5.77$, (c) $z = 6.24$, (d) $z = 6.34$, and (e) $z = 6.77$. Individual galaxies are color-coded by their line-of-sight velocity in km/s relative to the group’s mean redshift and scaled based on their stellar mass.}
    \label{fig:velocity_plot}
\end{figure*}
\noindent

This overdensity metric captures the relative enhancement in galaxy density compared to the cosmic mean at a given redshift. Uncertainties on $\rm \delta_{gal}$ are estimated by propagating Poisson errors on galaxy counts in both the group and the redshift bin. While this provides a first-order estimate of the statistical uncertainty, it does not account for systematic effects such as redshift uncertainties (which are small in this case due to the robust spectroscopic redshifts) or cosmic variance. In particular, if the survey field samples a region that is intrinsically over- or underdense, the resulting $\rm \delta_{gal}$ values may be biased. The overdensity values for each structure are listed in Table \ref{table:overdensities}.

\subsection{Inferring total halo masses}\label{sec:halo_mass}
To characterize the total mass scale of the identified overdensities, it is essential to estimate the masses of their associated dark matter halos which are expected to dominate their matter content. These estimates offer critical insight into the gravitational potentials of the structures, their likely evolutionary pathways into present-day galaxy clusters, and their role within the context of cosmic large-scale structure formation. We use three different methods for inferring the total halo masses, (1) a kinematic approach, (2) the empirical $\rm \textit{M}_{UV}-\textit{M}_{halo}$ relation from \cite{Mason_2023} for the sum of the individual $M_{\rm UV}$ magnitudes, and (3) a standard scaling from total stellar mass of the group members assuming a constant halo-to-stellar mass ratio. 

\subsubsection{Kinematic}
For each galaxy member in the overdensities, we calculate the line-of-sight (LOS) velocity as $v_{\text{LOS}} = c (z-z_{\text{cluster}})/(1 + z_{\text{cluster}})$, where $c$ is the speed of light, $z$ is the galaxy redshift, and $z_{\text{cluster}}$ is the cluster redshift. We define the latter as the mean redshift of the identified member galaxies, which serves as the reference frame for calculating the motion of individual galaxies relative to the group center \citep[e.g.,][]{Lee_2019}. The velocities of each galaxy member w.r.t. the cluster redshift can be seen for each overdensity in Figure \ref{fig:velocity_plot}.

To robustly estimate the velocity dispersion of the group, we employ the biweight midvariance as a statistical estimator, chosen for its resilience to outliers and observational uncertainties. The resulting LOS velocity dispersion, $\rm \sigma_{LOS}$, is subsequently used to infer the total halo mass. The adopted relation is given by:
\begin{align}
\rm \textit{M}_{halo} =\; & 1.2 \times 10^{15} \left( \frac{\sigma_{LOS}}{1000 \; km \; s^{-1}} \right)^{3} \left[ \frac{1}{\sqrt{ \Omega_{\Lambda} + \Omega_{m}(1+\textit{z}_{cluster})^{3} }} \right] \nonumber \\
& \times \left( \frac{\textit{H}_{0}}{100 \; km \; s^{-1} \; Mpc^{-1}} \right) \; M_{\odot},
\end{align}
where $H_{0}$ is the Hubble constant, and $\Omega_{\Lambda}$ and $\Omega_{m}$ are the cosmological density parameters \citep[see e.g. also][]{Lee_2019}. We infer total halo masses in the range $M_{\rm h}\sim 10^{11.5}-10^{12.5}\,M_\odot$ from this method. Although this assumes that the overdensities are approximately virialized, it provides a useful, luminosity-independent estimate of the total halo mass. In practice, however, many of these structures may still be in the process of gravitational collapse, with internal kinematics dominated by infall, mergers, or other non-equilibrium processes. Consequently, the measured velocity dispersions may be systematically biased high, potentially leading to overestimated halo masses. In addition the measured line-of-sight velocities are subject to systematic uncertainties in the spectroscopic redshifts, typically of the order $\sim 60\;\mathrm{km\;s^{-1}}$. These uncertainties artificially broaden the observed velocity distribution and are not explicitly accounted for in our analysis. As a result, the inferred velocity dispersions, and hence halo masses, may be mildly overestimated, particularly for systems with intrinsically low velocity dispersion. 

The associated uncertainty on the halo mass is derived via bootstrap resampling of the LOS velocities to estimate the error on $\sigma_{LOS}$, which is then propagated through the scaling relation assuming $M_{\mathrm{halo}} \propto \sigma_{LOS}^3$.

\subsubsection{$\rm \textit{M}_{UV}-\textit{M}_{halo}$ relation}
A second approach to estimating the dark matter halo mass of each overdensity relies on empirical relations between a galaxy's ultraviolet (UV) luminosity and the mass of its host halo. These relations are typically established via abundance matching techniques, which link the observed galaxy UV luminosity function to the theoretically predicted halo mass function from cosmological simulations. For this study, we adopt the relations from \citet{Mason_2023}, calibrated at $z \approx 6$, 7, and 8, which connects $\rm M_{UV}$ to the host halo mass via:
\begin{equation}
\rm \Phi(\textit{M}_{UV}) = \phi(\textit{M}_{h}) \left\vert \frac{\textit{dM}_{h}}{\textit{dM}_{UV}} \right\vert, 
\end{equation}
where the derivative is given by $\rm \textit{dM}{h}/\textit{dM}{UV} = \ln 10 , \textit{M}_{h} / 2.5$. We adopt the UV luminosities, $L_{\rm UV}$, derived by \citet{ALT_paper_2024} for individual ALT galaxies and sum them to obtain the total UV luminosity of each overdensity, which is then converted into an effective UV magnitude for the halo matching.
Using the adopted $M_{\mathrm{UV}} - M_{\mathrm{halo}}$ relations from \citet{Mason_2023}, we infer total halo masses of the order $M_{\rm h}\sim 10^{11.6}-10^{12.6}\,M_\odot$ (see Table~\ref{table:overdensities}). 
The associated uncertainty is primarily driven by the intrinsic scatter in the empirical relation, which is estimated to be on the order of $\sim$1.5 magnitudes in $M_{\mathrm{UV}}$.

\subsubsection{Constant halo-to-stellar mass ratio}
\begin{figure*}[h!]
    \centering
    \includegraphics[width=17cm]{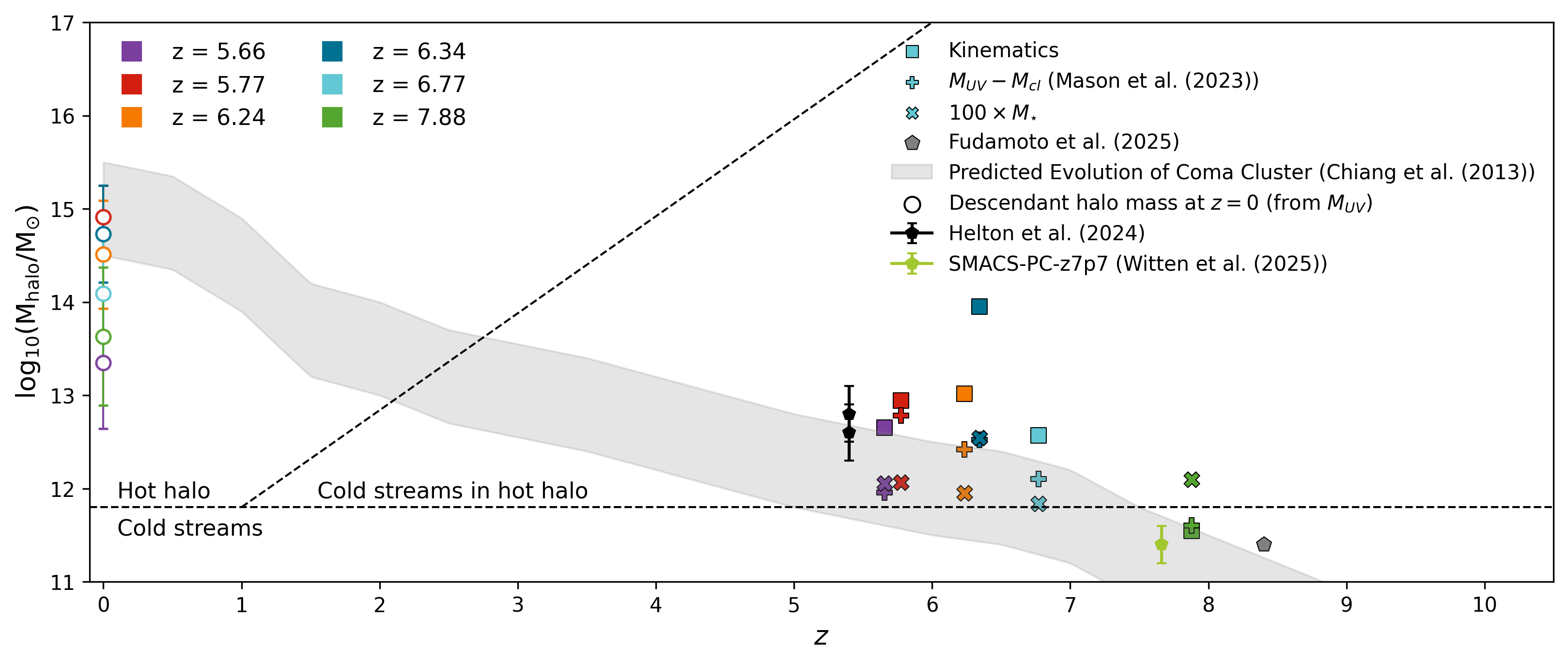}
    \caption[Dark matter halo mass versus redshift.]{Dark matter halo mass versus redshift. Each overdensity is color-coded following the notation of Fig.~\ref{fig:ODs_introfig}, with distinct symbols indicating the mass estimation method. Open color-coded circles at $z=0$ indicate the median descendant halo masses inferred from the UV-based halo masses, with error bars showing the standard deviation derived from tracing haloes in the IllustrisTNG simulations following the method described in \citet{Witten_2025_7_66_cluster}. The gray shaded region shows the expected evolution of a Coma-like cluster \citep{Chiang_2013} assuming constant evolution beyond $z \sim 7$. The black horizontal dashed line marks the shock stability threshold separating cold accretion from hot ICM formation \citep{Dekel_2006}, while the diagonal dashed line indicates the limit for penetrating cold flows. For comparison, overdensities from \citet{Helton_2024}, \citet{fudamoto_2025}, and \citet{Witten_2025_7_66_cluster} are shown with black, gray, and light green markers, respectively.}
    \label{fig:halo_mass_plot}
\end{figure*}

The third method estimates the dark matter halo mass of each overdensity based on the total stellar mass of its constituent galaxies. This approach assumes a fixed scaling relation between stellar and halo mass, motivated by empirical constraints and simulation-based studies \citep{Behroozi_2013}. Specifically, a constant halo-to-stellar mass ratio of 100 is adopted, reflecting the average scaling observed in abundance matching and halo occupation models. This yields total halo masses in the range $M_{\rm h}\sim 10^{11.2}-10^{12.1}\,M_\odot$  using this method. While this linear approximation captures the general trend, it does not account for variations in the halo-to-stellar mass ratio that may arise due to halo mass, redshift, or baryonic feedback processes, and is therefore used here only as an order-of-magnitude consistency check rather than a precise determination of halo mass.
The dark matter halo masses derived using the different methods are shown in Fig. \ref{fig:halo_mass_plot} and summarized in Table \ref{table:overdensities}, and are found to largely agree across all approaches.

\begin{table}
\centering
\caption{Properties of each identified overdensity analyzed in this work, including their respective redshifts, number of members ($N_{\mathrm{gal}}$), $\rm \delta_{gal}$, and estimated dark matter halo masses given as $\rm log_{10}(\textit{M}_{halo} / \textit{M}_{\odot})$.}
\begin{adjustbox}{width=9cm}
\begin{tabular}{l c c|c c c} 
 \hline
  $z$ & $N_{\mathrm{gal}}$ & $\rm \delta_{gal}$ & Kinematic & $\rm \textit{M}_{UV} - \textit{M}_{halo}$ & $100 \times M_{\star}$ \\ 
 \hline\hline 
  5.66 & 34 & 42.20 $\pm$ 8.89 & 12.65 $\pm$ 0.18 & 11.96 $\pm$ 0.6& 12.05$_{-1.74}^{+2.27}$ \\  [0.8ex]             
  5.77 & 59 & 29.44 $\pm$ 5.33 & 12.94 $\pm$ 0.25 & 12.78 $\pm$ 0.6& 12.06 $_{-1.81}^{+2.26}$ \\ [0.8ex]            
  6.24 & 33 & 27.56 $\pm$ 6.35 & 13.02 $\pm$ 0.12 & 12.42 $\pm$ 0.6& 11.95 $_{-1.81}^{+2.24}$ \\  [0.8ex]            
  6.34 & 43 & 32.41 $\pm$ 6.71 & 13.95 $\pm$ 0.39 & 12.52 $\pm$ 0.6& 12.54 $_{-1.79}^{+2.41}$ \\ [0.8ex]              
  6.77 & 21 & 41.00 $\pm$ 10.35 & 12.57 $\pm$ 0.46 & 12.10 $\pm$ 0.6& 11.84 $_{-1.66}^{+2.18}$ \\ [0.8ex]  
  7.88 & 4 & 7.00 $\pm$ 4.95 & 11.55 $\pm$ 0.53 & 11.60 $\pm$ 0.6& 12.10 $_{-1.74}^{+2.27}$ \\ [0.8ex]              
\hline
\end{tabular}
\end{adjustbox}
\label{table:overdensities}
\end{table}

\noindent
To place the inferred halo masses in a broader evolutionary context, we utilise our estimates of the halo masses of our sample of overdensities to identify similar haloes in the TNG300 \citep{Pillepich_2018, Nelson_2018, Naiman_2018, Marinacci_2018, Springel_2018} and TNG-Cluster \citep{Nelson_2024}. Following the method described in Sect. 5.1 of \citet{Witten_2025_7_66_cluster}, we trace comparable high-redshift haloes in these simulations down to $z=0$ to establish the distribution of descendant halo masses. We take the median and standard deviation of this distribution to establish the likely current-day halo mass and its associated uncertainty of each of our overdensities. 

The resulting $z = 0$ halo mass estimates provide an indication of the present-day descendants of the identified overdensities range between $\mathrm{log}_{10}(M_{\mathrm{halo}}/M_{\odot}) \sim 13.35 - 14.91$ and are shown in Fig.~\ref{fig:halo_mass_plot}. Given that galaxy clusters at $z=0$ are typically defined as systems with $M_{\rm halo} > 10^{14} \; M_{\odot}$, these results suggest that most of the identified overdensities are likely to evolve into present-day galaxy clusters, and can therefore be considered strong protocluster candidates.

\subsection{Modeling the \lya Absorption}\label{sec:dla_model}
\begin{figure}[h!]
    \centering
    \includegraphics[width=9cm]{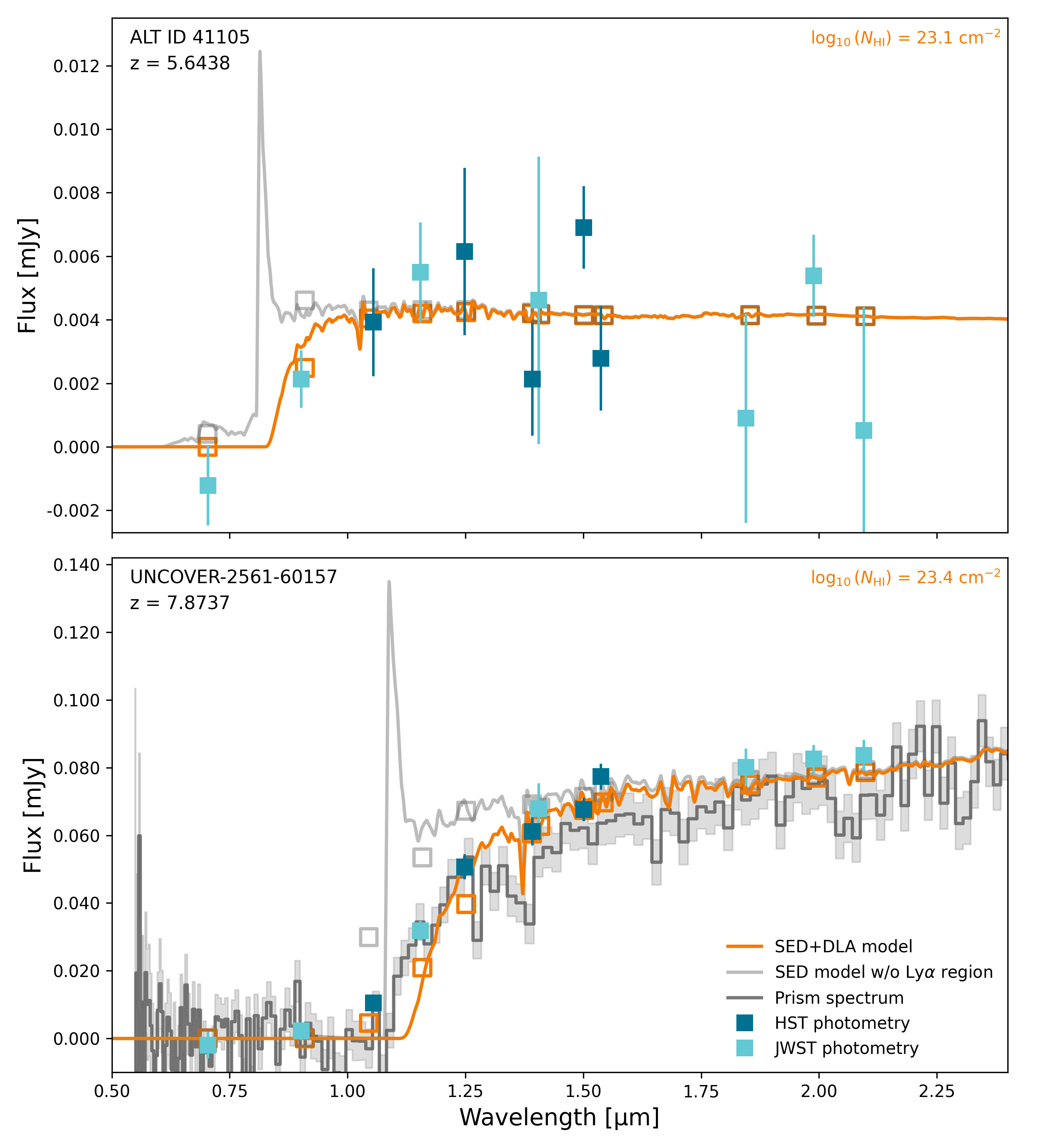}
    \caption[XX]{Example SED fits and photometric data for two galaxies in two of the overdense environments at $z = 5.64$ (top) and $z = 7.88$ (bottom). Filled squares with error bars show the HST (dark blue) and JWST/NIRCam (light blue) photometry. The dark gray step curve displays the NIRSpec/prism spectrum. The orange solid line shows the best-fitting SED model including a DLA (SED+DLA), while the semi-transparent grey curve indicates the corresponding intrinsic SED model without the Ly$\alpha$ region from \citet{ALT_paper_2024}. Open orange squares represent the model-predicted photometry obtained from the SED+DLA model, and open grey squares show the photometry predicted by the intrinsic SED. The inferred $N_{\hi}$  values are indicated in the upper right of each panel.} 
    \label{fig:prism_fit_example}
\end{figure}

To assess the presence of DLAs within each of the identified galaxies in the six most massive overdensities, we adopt the rest-frame UV spectral slopes ($\beta_{\rm UV}$) derived by \citet{ALT_paper_2024} as a baseline description of the intrinsic rest-frame UV continua. However, since the ALT SEDs do not include a treatment of the \lya\ region, they are unsuitable for directly constraining the neutral gas content of the galaxies. For this reason, we construct a complementary ``simple'' SED for each galaxy that explicitly models the continuum and absorption around \lya, while retaining the ALT SED redward of the line instead of interpolating a pure power-law continuum (see Fig. \ref{fig:prism_fit_example} for examples). In this framework, the continuum redward of $\lambda_{\rm Ly\alpha}(1+z_{\rm spec})+500\,\AA$ is described by a power law anchored to the ALT SED, $F_\lambda \propto \lambda^{\beta_{\rm UV}}$. The offset is chosen to avoid contamination from \lya\ and surrounding absorption features while remaining within the well-constrained photometric range. A Gaussian \lya emission line is included at the systemic redshift, with its flux tied to the observed H$\beta$ emission assuming case-B recombination ($R_{\rm Ly\alpha/H\beta}=32.7$ \citep{OsterbrockFerland}), such that any attenuation of \lya is attributed to absorption rather than an explicit escape fraction (i.e. assuming $f_{esc, \mathrm{Ly}\alpha} = 1$). A caveat of this approach is the redshift dependence to constrain $N_{\rm HI}$ from photometry, since the apparent strength and wavelength of the DLA depend on redshift and filter coverage. In our analysis, however, the combination of medium- and broad-band \textit{HST} and \jwst\ photometry provides dense coverage of the rest-frame UV, and the availability of precise spectroscopic redshifts removes any major degeneracy. As a result, the redshift-dependent uncertainties are not expected to significantly affect our conclusions, certainly for the internal cluster variations. 
For clarity, the different components contributing to the modeled \lya absorption are treated as follows:
\begin{itemize}
\item the intrinsic UV continuum, anchored to the ALT SED and parameterized by $\beta_{\rm UV}$,
\item intrinsic \lya emission at the systemic redshift, scaled from H$\beta$,
\item absorption by neutral hydrogen associated with the galaxy, modeled with a Voigt profile and parameterized by $N_{\rm HI}$,
\item additional absorption from the surrounding intergalactic medium, included via a standard damping-wing prescription.
\end{itemize}
The \lya\ absorption from neutral hydrogen in the interstellar medium (ISM) or the immediate surroundings of each galaxy is then modeled using a Voigt–Hjerting profile, following the analytical approximation of \citet{TepperGarcia06}:
\begin{equation}
\tau_{\rm ISM}(\lambda_{obs}) = C \; N_{\rm HI} \; a \; H(a,x),
\end{equation}
where $C$ is the absorption constant, $a$ is the damping parameter, $N_{\rm HI}$ the neutral hydrogen column density, and $H(a,x)$ the Voigt-Hjerting function \citep[see e.g.,][]{Heintz_2024_exDLA}. We also account for the absorption expected from the surrounding IGM. As a baseline, we adopt the Gunn-Peterson optical depth, with the extended damping wing modeled following the formalism of \citet{Miralda_Escude_2000} and \citet{Totani_2006}:
\begin{multline}
\tau_{\rm IGM}(\lambda_{\rm obs}, z) = \frac{x_{\rm HI} R_{\alpha} \tau_{\rm GP}(z_{\rm gal})}{\pi}
\left( \frac{1+z_{\rm abs}}{1+z_{\rm gal}} \right)^{3/2} \\
\times \Big[ I\left( \frac{1+z_{\rm IGM,u}}{1+z_{\rm abs}} \right) - I\left( \frac{1+z_{\rm IGM,l}}{1+z_{\rm abs}} \right) \Big],
\end{multline}
where $x_{\rm HI}$ is the average neutral-to-total hydrogen fraction of the IGM, set artificially small in order to ensure that the absorption effectively traces the total \hi column density at each galaxy, $z_{\rm gal}$ is the galaxy’s systemic redshift, and $R_{\alpha} = \Lambda_{\alpha}\lambda_{\rm Ly\alpha}/(4\pi c) = 2.02 \times 10^{-8}$ is a dimensionless factor depending on the \lya damping constant $\Lambda_{\alpha}$ and rest wavelength $\lambda_{\rm Ly\alpha}$. $I(x)$ is the analytic integral of the \lya damping-wing opacity, which accounts for the cumulative absorption by neutral hydrogen along the line of sight given as: 
\begin{equation}
    I(x) = \frac{x^{9/2}}{1-x} + \frac{9}{7}x^{7/2} + \frac{9}{5}x^{5/2} + 3x^{3/2} + 9x^{1/2} - \frac{9}{2}\mathrm{ln}\left( \frac{1 + x^{1/2}}{1 - x^{1/2}} \right),
\end{equation}
which is mostly a good approximation assuming $(z_{\rm abs} - z_{\rm IGM,u}) \gg R_{\alpha}(1 + z_{\rm abs})$ \citep[see][]{Totani_2006}. For each galaxy we set the upper bound at the galaxy redshift ($z_{\rm IGM,u} = z_{\rm gal}$) and the lower bound to $z_{\rm IGM,l} = 5.0$. The Gunn-Peterson optical depth is given by
\begin{equation}
    \tau_{GP}(z) = 1.8 \times 10^{5}h^{-1}(\Omega_{\rm DM,0})^{-1/2}x_{\rm HI}\left( \frac{\Omega_{\rm m,0}h^{2}}{0.02}\right) \left( \frac{1+z}{7} \right)^{3/2},
    \label{eq:GP_1}
\end{equation}
where $h = H_{0}/(100\,{\rm km\,s^{-1}\,Mpc^{-1}})$ is the present-day dimensionless Hubble parameter, and $\Omega_{\rm DM,0}$ and $\Omega_{\rm m,0}$ are the current dark matter and matter density parameters \citep{Fan_2006, Heintz_2024_exDLA}. For this analysis, we assume that the DLA arises from gas in close proximity to the galaxy itself and therefore fix the absorber redshift to the galaxy redshift ($z_{\rm abs} = z_{\rm gal}$). We caution that we are in most cases unable to constrain absorption redshifts \citep[as done by e.g.,][]{Heintz_2026_Nat, Terp_2024}, since we are relying on photometric data alone in the rest-frame UV part of the spectrum. We will evaluate this method further in Sect.~\ref{sec:neutral_gas}. 

\section{Galaxy properties in overdense environments}\label{sec:gal_properties}
The overdensities identified in Sect.~\ref{sec:ODs} provide a unique opportunity to investigate how galaxy properties are shaped by their large-scale environments during the EoR. In this section, we compare the stellar population and ISM properties of galaxies residing in the six FoF–selected overdensities to those of field galaxies drawn from the ALT survey at a similar redshift. To investigate whether the physical properties of galaxies residing in overdense environments differ from those of the field population, we focus on comparing the following main characteristics; UV magnitudes, stellar masses, UV continuum slopes, metallicities, and star formation rates for each overdensity to the full ALT field-selected galaxy benchmark sample. 
We consider the full redshift-space that covers H$\beta$ and [O\,{\sc iii}] at $z\approx 5.5-7$ to ensure the statistical power needed to characterize the underlying field population, since the number of field galaxies within the narrow redshift ranges spanned by individual overdensities is too small to establish reliable reference distributions. This approach allows us to identify potential environmental trends while avoiding noise driven by low-number statistics.



\subsection{Stellar masses}\label{sec:stellar_mass}
\begin{figure*}[h!]
    \centering
    \includegraphics[width=17cm]{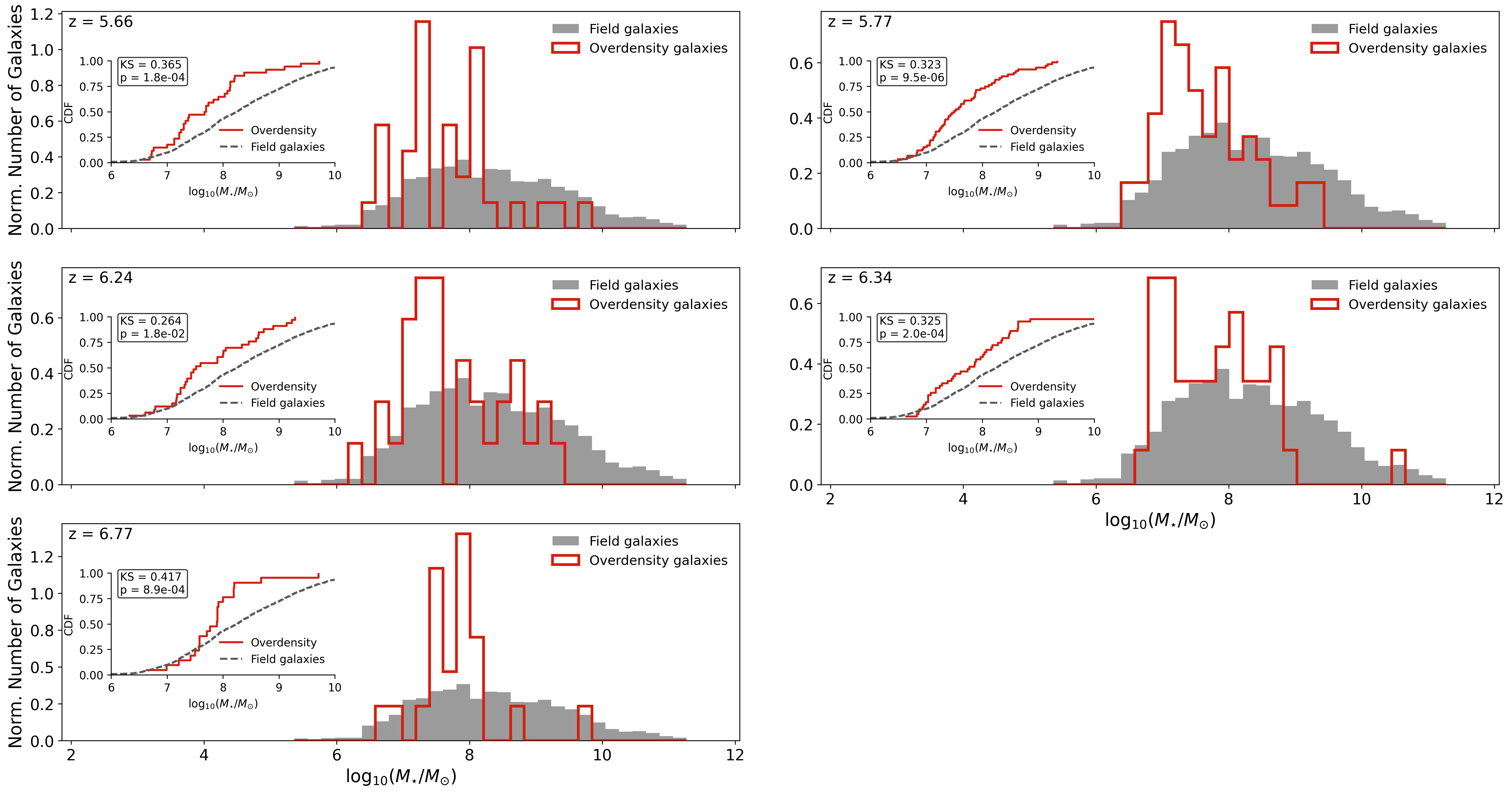}
    \caption[XX]{Normalized distributions of stellar masses ($\log_{10}(M_{\star}/M_{\odot})$) for overdensity galaxies (red) and field galaxies (gray) in each of the five largest overdensities. Insets show the corresponding empirical cumulative distribution function (ECDFs) along with the KS statistic and $p$-value.}
    \label{fig:stellar_hist}
\end{figure*}

We first compare the stellar masses of galaxies in each overdensity to those of the full ALT field population in Fig. \ref{fig:stellar_hist}. Across all five structures, the overdensity members span a broad range of $\log_{10}(M_\star/M_\odot) = 6.5-9.5$. We see clear evidence for their distributions being shifted towards lower masses compared to the general field population, with $p$-values $p<0.01$. The stellar mass distribution of the field galaxies exhibits a peak around $\log_{10}(M_\star/M_\odot)\sim 8-9$ and extends to $\gtrsim 10^{10.5},M_\odot$. None of the identified overdensity galaxies exhibit such high stellar masses. 

At first glance, this result may appear counter-intuitive, as galaxies residing in massive dark matter overdensities are expected to experience earlier collapse and enhanced gas accretion, leading to potentially rapid stellar mass growth. First, while the inferred halo masses of the overdensities are already substantial, the overdensities likely correspond to an early evolutionary phase in which baryonic assembly is still ongoing. In such environments, efficient gas accretion does not necessarily translate immediately into high stellar masses, particularly if star formation is `bursty' or temporarily suppressed. In addition, selection effects may also play a role. The ALT sample is fundamentally emission-line selected, relying on the detection of rest-frame optical lines such as [O\,{\sc iii}] and  H$\beta$. This imposes an effective SFR threshold, such that massive galaxies with low (or declining) star formation activity may fall below the detection limits and therefore remain absent from the sample. As a result, the observed stellar mass distributions likely trace the actively star-forming subset of galaxies within the overdensities, rather than the full underlying population. 


\noindent


\subsection{Rest-frame UV continuum slopes, UV luminosities, and star-formation rates} \label{sec:uv_properties}
The rest-frame UV properties, $\beta_{\rm UV}$ and $M_{\rm UV}$, together with the Balmer breaks $D_{\rm B}$ and SFRs derived from H$\beta$, probe the star formation activity on different timescales and provide insights into the typical ages of the stellar populations of the overdensity members. We find no statistically significant difference ($p>0.1$) in terms of the UV brightness (probing SFRs on $\sim 100$\,Myr timescales) or their more instantaneous SFR ($\sim 10$\,Myr) derived from H$\beta$ when comparing the cluster-members to field-selected galaxies at similar redshifts. 

In contrast, we find significant evidence for the rest-frame UV continuum slopes of the galaxy overdensity members being, on average, bluer than those of the field population. The mean (and median) spectral slopes for the cluster-members are $\beta_{\rm UV,mean} = -2.12$ ($\beta_{\rm UV,med} = -2.21$), where the field-selected galaxies are significantly redder with $\beta_{\rm UV,mean} = -1.67$ ($\beta_{\rm UV,med} = -1.93$), with $p<0.01$ for the five clusters at $z=5.66-6.77$. 

\begin{figure*}[h!]
    \centering
    \includegraphics[width=17cm]{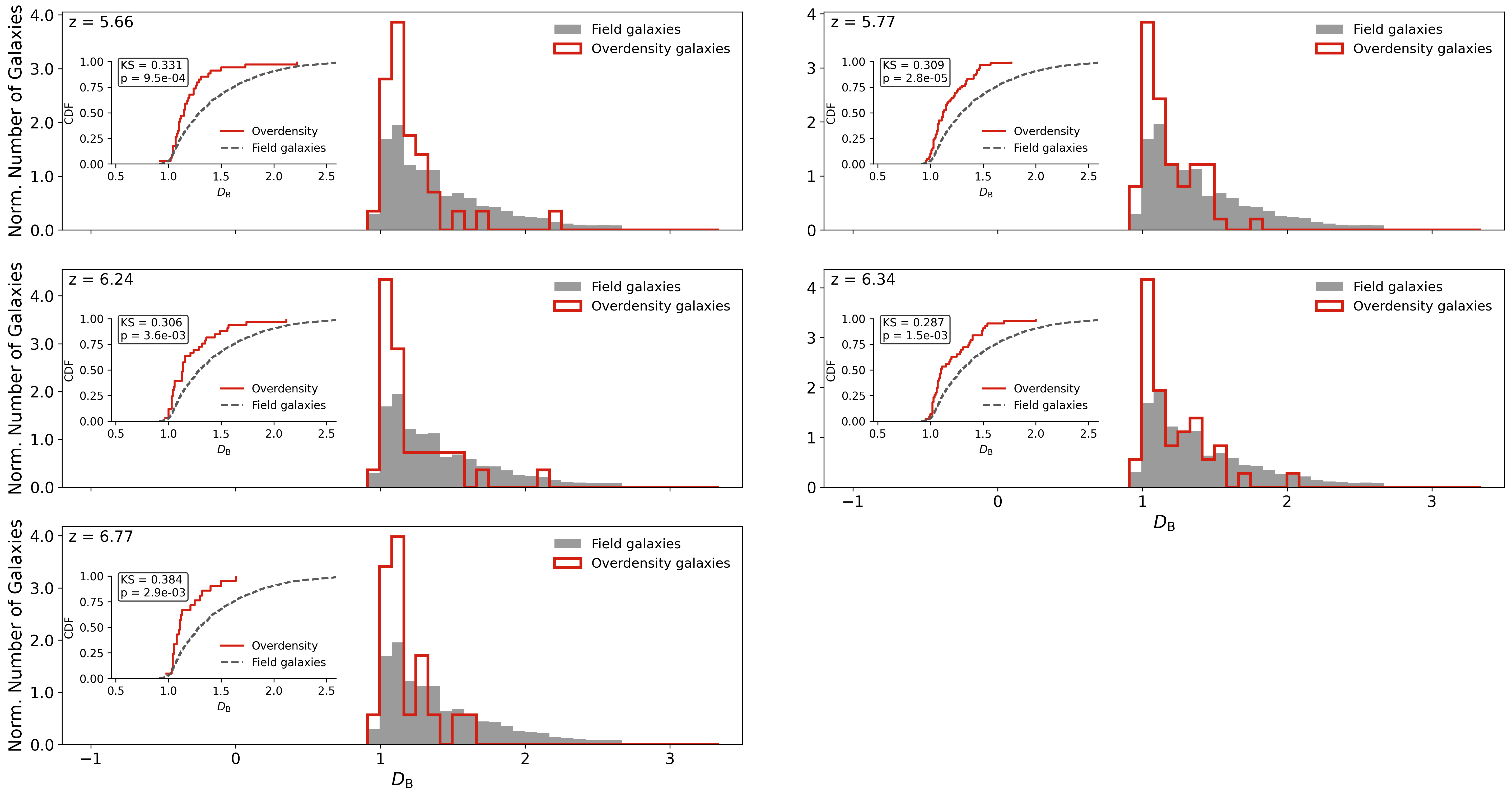}
    \caption[XX]{Normalized distributions of Balmer break strength ($D_{\mathrm{B}}$) for the overdensity and field galaxies. The colors of the two distributions and the insets follow the same notation as in Fig.~\ref{fig:stellar_hist}. }
    \label{fig:balmer_B_hist}
\end{figure*}

This interpretation is further supported by the distribution of Balmer break strengths, $D_{\rm B}$, as shown in Fig.~\ref{fig:balmer_B_hist}. The overdensity galaxies exhibit weaker Balmer breaks than the field population at fixed redshift, indicating younger luminosity-weighted stellar ages and a reduced contribution from older stellar populations. Taken together, the bluer continua and weaker Balmer breaks suggest that galaxies in overdense environments are dominated by relatively young stellar populations, despite having comparable UV luminosities to field galaxies. Combined with their systematically lower stellar masses, this implies that star formation in these particular galaxies do not seem to accelerate their stellar evolution, as also noted for another galaxy overdensity at $z=7.66$ \citep{Witten_2025_7_66_cluster}. 


\subsection{Gas phase metallicities}\label{sec:metal}
\begin{figure*}[h!]
    \centering
    \includegraphics[width=17cm]{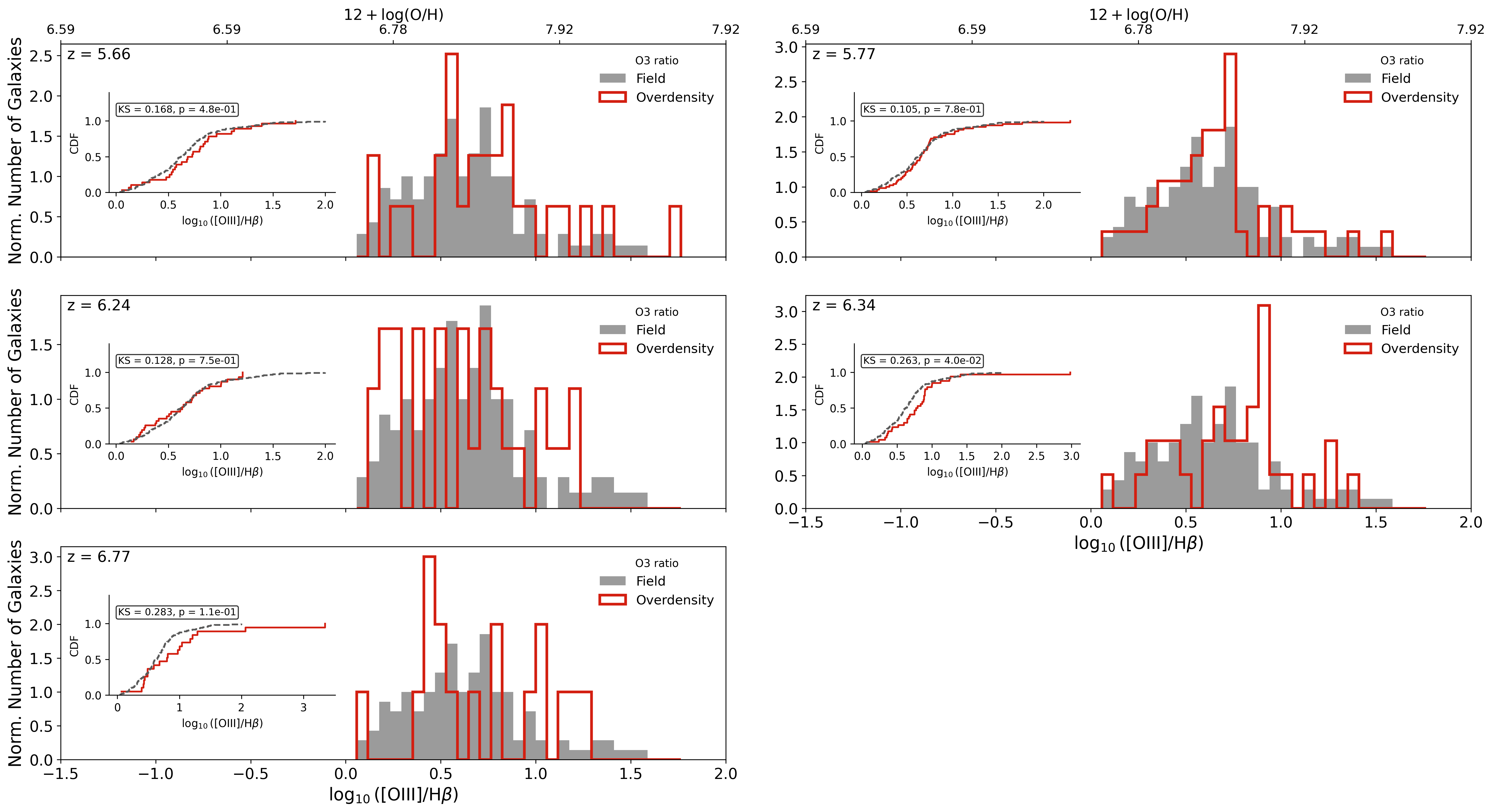}
    \caption[XX]{Normalized distributions of the gas-phase metallicity proxy $\log_{10}(\mathrm{[O,III]}/\mathrm{H}\beta)$ for overdensity galaxies and field galaxies. The colors of the two distributions and the insets follow the same notation as in Fig.~\ref{fig:stellar_hist}. The upper axes indicate the corresponding oxygen abundances, $12+\log(\mathrm{O/H})$ derived from the strong-line ratio from \cite{Sanders24}, for reference.}
    \label{fig:met_hist}
\end{figure*}

For lower-mass galaxies with, on average, younger stellar populations as inferred for the majority of the overdensity systems, we expect the metal abundance to be equally lower. In Figure~\ref{fig:met_hist}, we compare the metallicity distributions of the five clusters to the field-galaxy benchmark sample, using the [O\,{\sc iii}]\,$\lambda 5007$/H$\beta$ (O3) ratio as a proxy for the gas-phase metallicity $Z$, interpreted through the recent \jwst-based calibrations from \citet{Sanders24}. In this framework, higher O3 values correspond to lower metallicities, while lower O3 values indicate more metal-rich gas.


Contrary to the expectations, we find that the metallicities are systematically higher for the overdensity members of the three systems at $z>6$, though not at high significance ($p=0.01-0.2$). The three lower-redshift systems do not show any evidence for enhanced metallicities compared to the field-average. This does still imply, however, in combination with the systematically lower stellar masses across all five systems, that the chemical enrichment of the overdensity-members are generally higher at a given stellar mass. This could indicate a potential enhanced star-formation efficiency during an earlier time in the star-formation history of the galaxies \citep{Ellison_2008}, or potentially a higher fraction of high-mass stars yielding a higher chemical yield, as also supported by the, on average, bluer spectral slopes. 

\section{Neutral gas abundances in reionization-era galaxy overdensities}\label{sec:neutral_gas}
The blind, rest-frame optical selection delivered by the ALT survey strategy allow us to obtain an unbiased census of the fraction of strong Lyman-$\alpha$ emitters (LAEs) and galaxy DLAs in proto-cluster environments in the reionization era. This is particularly important to probe the overall baryonic matter budget of these systems, and given that past surveys have mainly relied on bright LAEs as signposts of galaxy overdensities, both pre- \citep[e.g.,][]{Castellano_2018,Leonova_2022} and post-\jwst\ \citep[e.g.,][]{Chen_2025,Witstok_2025}. The neutral gas content of galaxies in the overdensities derived from the Voigt-profile modelling of Ly$\alpha$ absorption as described in Sect. \ref{sec:dla_model} further allow us to model the tomography of \hi\ in these overdense regions. This has previously only been done in the outskirts of a similar-mass proto-cluster at $z>5$ \citep{Heintz_2026_Nat}, likely tracing the filamentary, accreting cold gas stream and not the central components. 

As a first validation check for the robustness of photometric DLA modelling, we search DJA for \jwst/NIRSpec Prism spectra of any of the identified cluster members. We show two examples in Fig.~\ref{fig:prism_fit_example}, comparing the intrinsic SED model with the best-fit DLA model and for one of the galaxies the observed Prism spectrum. For both galaxies with Prism spectroscopy, the $N_{\mathrm{HI}}$ values inferred from the full spectral shape are consistent with those obtained from the photometry alone. The close agreement between the SED+DLA model and the broadband fluxes shows that even in the absence of spectroscopy, the photometric data contain enough information to meaningfully constrain strong Ly$\alpha$ absorption.

\begin{figure*}[h!]
    \centering
    \includegraphics[width=18cm]{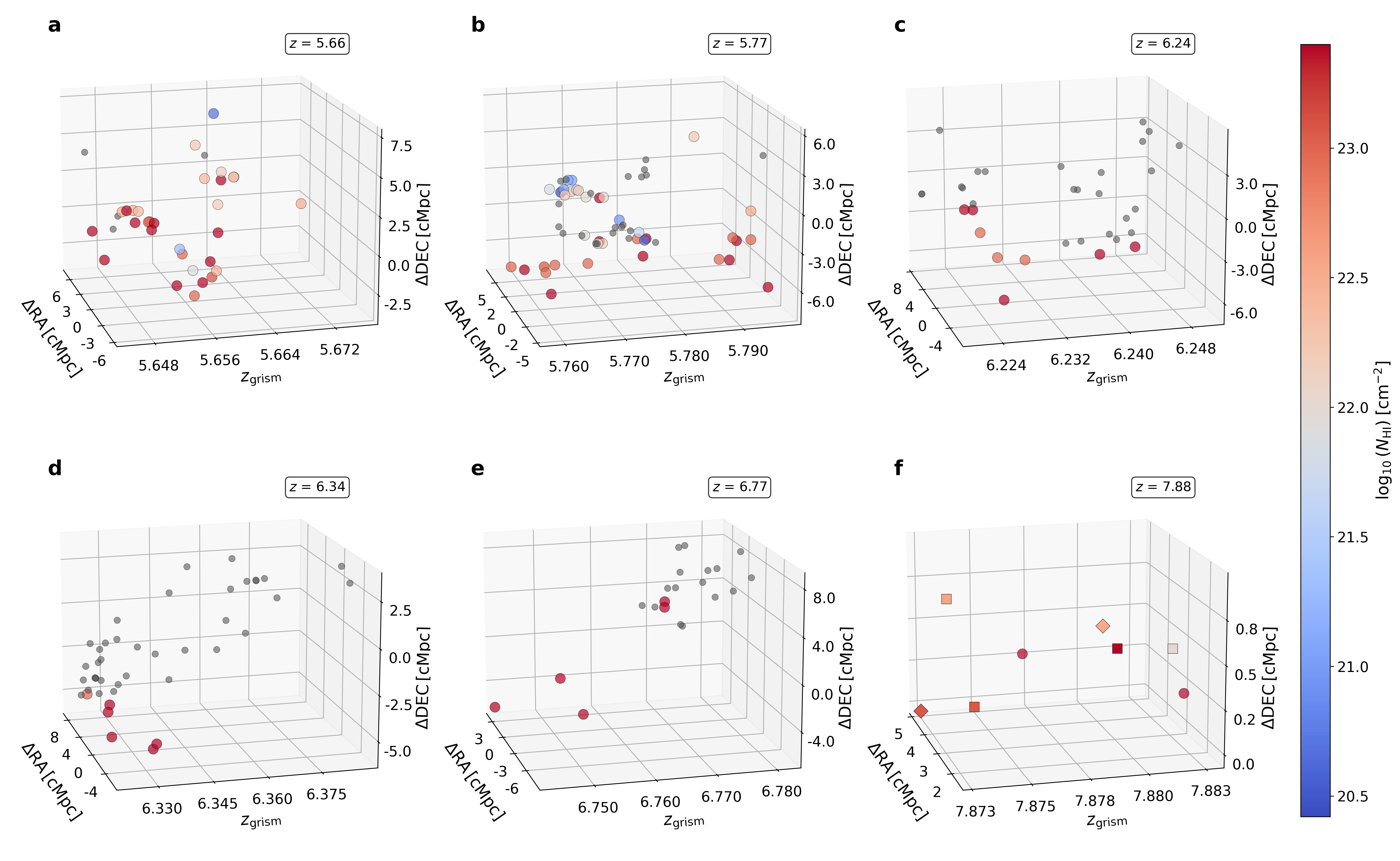}
    \caption[XX]{Three-dimensional spatial distribution of galaxies in the five identified groups as well as the overdensity at $z = 7.88$, shown in comoving coordinates $\Delta$RA, $\Delta$DEC, and grism redshift $z$. Each panel corresponds to one overdensity. Points are colored by the best-fit neutral hydrogen column density $\log_{10}(N_{\mathrm{HI}})$: galaxies with $\log_{10}(N_{\mathrm{HI}}) > 20$ use a shared color scale (right), while galaxies with $\log_{10}(N_{\mathrm{HI}}) \leq 20$ are shown in grey. A common colorbar is used for all high-$N_{\mathrm{HI}}$ values to enable direct comparison across groups. In panel (f), the two galaxies observed with the NIRSpec prism and associated with ALT sources are highlighted with diamond symbols and colored according to their prism-based $N_{\mathrm{HI}}$ measurements. Square symbols denote additional galaxies with prism spectra that are not part of the ALT survey also colored according to their prism-based $N_{\mathrm{HI}}$ measurements.}
    \label{fig:n_HI_all_3D}
\end{figure*}

The derived column densities span $\log_{10}(N_{\mathrm{HI}} / \mathrm{cm}^{-2}) \approx 18-23$. Since most models are derived from photometry alone we are not able to identify LAEs reliably, but note that the preferred models with column densities $N_{\mathrm{HI}} \lesssim 10^{20}$ cm$^{-2}$ are likely indiscernible from LAEs. We also do not explicitly account for potential AGN contamination in the modeling, which may introduce additional uncertainty in individual cases. In the following, we thus only consider galaxies with $N_{\mathrm{HI}} \geq 10^{20}$ cm$^{-2}$ for the \hi\ tomography. We note that the inferred LAE fraction varies substantially between overdensities, ranging from $\sim 0$ in the $z = 7.88$ overdensity to $\sim 0.8$.
The spatial distributions of $N_{\mathrm{HI}}$ for each overdensity are shown in Figure \ref{fig:n_HI_all_3D}. Galaxies with extreme column densities, approaching or exceeding $N_{\mathrm{HI}} \gtrsim 10^{22}$ cm$^{-2}$, are found in all of the overdensities. In particular, every single galaxy in the highest-redshift overdensity at $z = 7.88$ exhibits a high column density as also noted previously \citep{Chen_2024}, and consistent with a recent study of a massive galaxy overdensity at $z=7.66$ \citep{Witten_2025_7_66_cluster}. 

\noindent

Across the six overdensities, the distribution of $N_{\mathrm{HI}}$ values shows substantial internal variation, with galaxies in the same structure often spanning several orders of magnitude in column density. For example, in the overdensities at $z = 5.66$, $5.77$, and $6.24$, systems with $\log_{10}(N_{\mathrm{HI}} / \mathrm{cm}^{-2}) > 21.5$ are scattered among galaxies with much lower column densities, indicating that the presence of strong Ly$\alpha$ absorption is not confined to a specific spatial subregion of the structure. Instead, high-column systems appear throughout the full range of comoving offsets in RA, Dec, and redshift, suggesting that large neutral gas reservoirs are widespread within these environments rather than associated with a single core or filament.

The overdensities at $z = 6.34$ and $6.77$ display a similar pattern. Although they contain fewer extreme systems, the galaxies with elevated column densities do not cluster spatially. This lack of spatial segregation implies that the mechanisms driving the large neutral columns -- whether internal ISM geometry, local ionization conditions, or the presence of patchy neutral gas in the surrounding medium -- operate across the entire overdensity rather than in a single localized region.

\begin{figure*}[h!]
    \centering
    \includegraphics[width=17cm]{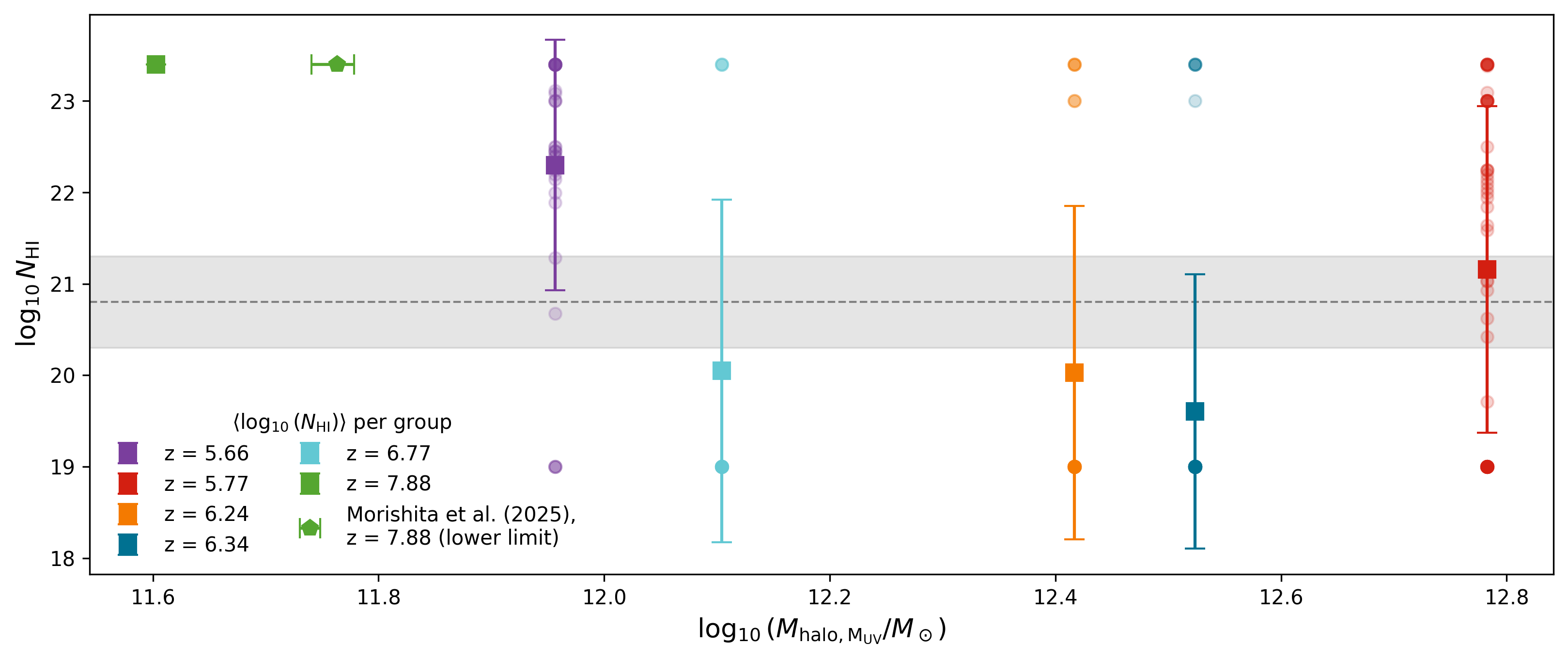}
    \caption[XX]{Mean neutral hydrogen column density of each overdensity as a function of its $M_{UV}$-inferred halo mass. Each point corresponds to one overdensity and is colour-coded by redshift. Error bars indicate the dispersion in $N_{\mathrm{HI}}$ among member galaxies. The dashed line and grey-shaded band represent the median $N_{\rm HI}$ of field galaxies at $z>5$ from \citet{Mason_2025} and a 0.5 dex scatter. The green hexagon represent the estimated $M_{UV}$-inferred halo mass of the $z = 7.88$ overdensity by \cite{Morishita_2025b}.}
    \label{fig:halo_mass_vs_NHI}
\end{figure*}

Figure \ref{fig:halo_mass_vs_NHI} places the overdensities in the context of field galaxies at $z>5$ by comparison with the median neutral hydrogen column densities inferred from \citet{Mason_2025}. While the field population exhibits a relatively modest median 
$N_{\rm HI}$ with substantial intrinsic scatter, several of the ALT overdensities—most notably the highest-redshift structure at $z=7.88$ lie systematically above the field relation. This indicates that, at fixed inferred halo mass, galaxies in some overdense environments host larger neutral gas columns than are typically observed in the field. In contrast, other overdensities overlap with, or fall within, the scatter of the field population, suggesting that enhanced neutral gas content is not a universal property of overdense regions at these epochs.

The $z = 7.88$ overdensity stands out as the most coherent case. All four galaxies exhibit high column densities, and the two sources with NIRSpec Prism spectra that are spatially well matched to the photometric positions (highlighted with diamond symbols in Fig.~\ref{fig:n_HI_all_3D}) show some of the strongest and most clearly defined damping wings in the sample. Although this overdensity spans a relatively compact comoving volume, the uniformity of its high $N_{\mathrm{HI}}$ values suggests that the entire structure may reside within a particularly dense and highly neutral region of the cosmic web.

A similar diversity in neutral gas properties has also been observed in other high-redshift overdensities. For example, \cite{Witten_2025_7_66_cluster} report that the $z=7.66$ protocluster SMACS-PC-z7p7 hosts galaxies with neutral hydrogen columns comparable to the surrounding field population, suggesting that high $N_{\mathrm{HI}}$ is not a universal property of early overdense environments. This supports our finding that only some of the ALT overdensities -- most notably the structure at $z=7.88$ -- show uniformly elevated column densities, while others span several dex in $N_{\mathrm{HI}}$. A tentative pattern can be seen when comparing the average neutral gas content with the inferred halo masses of the overdensities (Fig.~\ref{fig:halo_mass_vs_NHI}). Overdensities at higher redshifts tend to exhibit elevated mean $N_{\mathrm{HI}}$ values, while systems at lower redshift span a wider range of neutral gas content. A similar level of scatter is seen when comparing $N_{\mathrm{HI}}$ with $M_{UV}$-inferred halo mass: overdensities with comparable inferred masses can host noticeably different column densities, demonstrating that the neutral gas content does not scale monotonically with halo mass inferred from UV luminosity. Such variations are expected, given also that the $N_{\rm HI}$ derived for each galaxy is sightline dependent to that particular system as well \citep[see e.g.][]{Gelli_2025}. Even so, the overall distribution hints that the most massive and highest-redshift overdensities may be more likely to host uniformly large neutral gas reservoirs, whereas the remaining structures show a wider diversity in $N_{\mathrm{HI}}$ values.

There are, however, clear exceptions: for instance, the $z=5.66$ overdensity has a higher mean $N_{\mathrm{HI}}$ than the slightly more massive $z=6.24$ structure, illustrating that neutral gas content does not increase monotonically with halo mass. 

This interpretation is broadly consistent with the findings of \citet{Witten_2025_7_66_cluster}, who report that galaxies in the $z=7.66$ SMACS-PC-z7p7 protocluster show no clear environmental signatures and argue that stronger environmental effects may only emerge once overdensities exceed a halo-mass threshold of $\log_{10}(M_{\mathrm{halo}}/M_{\odot}) \approx 11.5$. Although our sample is limited, the elevated and spatially coherent neutral gas content in the most massive overdensity, together with the substantial scatter across the others, supports a picture in which both halo mass and evolutionary stage modulate the neutral gas properties of galaxies in early overdense environments, but with considerable variation between individual systems.

\balance
\section{Conclusions}\label{sec:conc}
We have presented a blind, rest-frame optical search for and characterization of galaxy overdensities during the EoR at $z \sim 5.5 - 7$, based on \jwst/NIRCam grism spectroscopy from the {\tt ALT} survey. By applying a physically motivated FoF algorithm to a spectroscopically confirmed sample, we identified five significant overdensities, spanning redshifts $z = 5.66$ to $6.77$. Independent halo mass estimates based on galaxy kinematics, UV luminosities, and total stellar mass consistently imply total halo masses $M_{\mathrm{halo}} \sim 10^{11} - 10^{14} \; M_{\odot}$, placing these systems among some of the most massive known structures at these epochs. 

We compared the physical properties of galaxies residing within the overdensities to a control sample of field galaxies at comparable redshifts drawn from the ALT survey. Across all five robustly identified structures, overdensity members span a broad range of stellar masses, but are, on average, systematically less massive than field galaxies at similar redshifts, while exhibiting comparable UV luminosities and SFRs derived from $H\beta$. Despite this, the overdensity members show systematically bluer UV continua and weaker Balmer breaks, indicating younger luminosity-weighted stellar populations. Together with their elevated metallicities at fixed stellar mass, this suggests that star formation in these environments does not proceed in a simply accelerated manner, but may instead reflect more efficient or earlier episodes of star formation compared to the field.

We further examined the chemical enrichment of the overdensity galaxies using gas-phase metallicites inferred from rest-frame optical emission line rations. At $z > 6$, we find tentative evidence that galaxies in overdense environments are more chemically enriched compared to their field counterparts, while no such trend is observed for the lower-redshift overdensities. Although the statistically significance remain modest, this result suggests that environmental effects may influence early chemical enrichment, potentially through enhanced star-formation efficiency or differences in stellar populations at earlier times. 

Using detailed modeling of \lya\ absorption, we traced the neutral hydrogen content of galaxies within each overdensity. The inferred column densities span $\mathrm{log}_{10}(N_{\mathrm{HI}}) \approx 18 - 23$, with several systems reaching the DLA regime. The spatial distribution of high-$N_{\mathrm{HI}}$ systems within the overdensities does not reveal strong evidence for segregation within individual structures. Galaxies with extreme column densities (DLAs) are found throughout the full spatial and redshift extent of the overdensities, suggesting that large neutral gas reservoirs are widespread rather than confined to specific cores or filaments. The highest redshift overdensity at $z = 7.88$ stands out in this regard, as all member galaxies exhibit high \hi\ column densities, consistent with the structure residing within a particularly dense and neutral region of the cosmic web. 

Taken together, our results demonstrate that massive galaxy overdensities during the reionization era do not necessarily host galaxies that are more massive and evolved than field galaxies. Instead, these systems appear to be characterized by relatively young, low-mass galaxies embedded within large and spatially extended reservoirs of neutral gas. This coexistence of typical stellar masses and metallicities with extreme $N_{\mathrm{HI}}$ implies that overdensities at $z > 5$ remain deeply embedded in gas-rich, partially neutral environments of the cosmic web. These reservoirs may correspond to filaments funneling cold gas into forming protoclusters, supplying fuel for future star formation while simultaneously delaying the local progress of reionization. 

These findings highlight the complex interplay between environment, gas accretion, and star formation in shaping galaxy evolution during the reionization era. In particular, they suggest that overdense regions at high redshift trace sites of enhanced gas supply rather than uniformly accelerated galaxy growth. Future deep, wide-area \jwst\ surveys and follow-up spectroscopy will be essential to determine how common such gas-rich overdensities are, and to establish how their baryonic content and star-formation activity evolve as these early structures mature into present-day galaxy clusters.



\begin{acknowledgements}
We would like to thank all the observers world-wide for their substantial effort in securing all the public \jwst\ data that were essential for this work. 

The Cosmic Dawn Center (DAWN) is funded by the Danish National Research Foundation under grant DNRF140.
KEH acknowledges support from the Independent Research Fund Denmark (DFF) under grant 5251-00009B and co-funding by the European Union (ERC, HEAVYMETAL, 101071865). Views and opinions expressed are, however, those of the authors only and do not necessarily reflect those of the European Union or the European Research Council. Neither the European Union nor the granting authority can be held responsible for them. 
This work is based in part on observations made with the NASA/ESA/CSA James Webb Space Telescope. The data were obtained from the Mikulski Archive for Space Telescopes (MAST) at the Space Telescope Science Institute, which is operated by the Association of Universities for Research in Astronomy, Inc., under NASA contract NAS 5-03127 for JWST. 

We used the following software for this work: \texttt{Python}, and the scientific Python ecosystem, in particular \texttt{NumPy} \citep{Harris_2020}, \texttt{SciPy} (including \texttt{cKDTree}) \citep{Virtanen2021}, \texttt{Matplotlib} \citep{Hunter_2007}, and \texttt{Astropy} \citep{Astropy_2013,Astropy_2018,Astropy_2022}. 
\end{acknowledgements}

\bibliographystyle{aa}
\bibliography{ref}

\end{document}